\documentclass[aps,prl,reprint,superscriptaddress,showpacs]{revtex4-1}

\usepackage{amssymb,amsmath,amssymb,amsfonts,amsthm,stmaryrd,mathrsfs,physics,bbm,mathtools}
\usepackage{graphicx}
\usepackage[T1]{fontenc}
\usepackage{xcolor}
\usepackage{bm} 
\usepackage{hyperref} % for HyperTeX cross-referencing

\newcommand{\p}{\hat{\mathbf{P}}}
\newcommand{\leftscript}[2]{\prescript{#2\csname prescriptcorrection\detokenize{#1}\endcsname {\mkern-8mu\relax}}{}{#1}}

\def\bee{\begin{eqnarray}}
\def\eee{\end{eqnarray}}

\newcommand{\ce}{\mathcal E}

\newcommand{\cg}{\mathcal G}
\newcommand{\ch}{\mathcal H}
\newcommand{\ci}{\mathcal I}

\newcommand{\ct}{\mathcal T}

\newcommand{\fh}{\mathfrak{h}}

\renewcommand{\b}{\beta}
\newcommand{\ga}{\gamma}
\newcommand{\G}{\Gamma}

\newcommand{\eps}{\varepsilon}

\newcommand{\sig}{\sigma}
\newcommand{\Sig}{\Sigma}

\renewcommand{\o}{\omega}
\renewcommand{\O}{\Omega}
\renewcommand{\t}{\tau}

\newcommand{\rmd}{\mathrm d}

\newcommand{\lt}{\left}
\newcommand{\rt}{\right}

\newcommand{\lag}{\left\langle}
\newcommand{\rag}{\right\rangle}

\begin{document}

\title{Fermions on Quantum Geometry and Resolution of Doubling Problem}

\author{Cong Zhang}
\email{czhang(AT)fuw.edu.pl}
\affiliation{Faculty of Physics, University of Warsaw, Pasteura 5, 02-093 Warsaw, Poland}

\author{Hongguang Liu}
\email{hongguang.liu(AT)gravity.fau.de}
\affiliation{Institut f\"ur Quantengravitation, Friedrich-Alexander-Universit\"at Erlangen-N\"urnberg, Staudtstr. 7/B2, 91058 Erlangen, Germany}

\author{Muxin Han}
\email{hanm(AT)fau.edu}
\affiliation{Department of Physics, Florida Atlantic University, 777 Glades Road, Boca Raton, FL 33431, USA}
\affiliation{Institut f\"ur Quantengravitation, Friedrich-Alexander-Universit\"at Erlangen-N\"urnberg, Staudtstr. 7/B2, 91058 Erlangen, Germany}

\begin{abstract}

The fermion doubling problem has an important impact on quantum gravity, by revealing the tension between fermion and the fundamental discreteness of quantum spacetime. In this work, we discover that in Loop Quantum Gravity, the quantum geometry involving superposition of states associated with lattice refinements provides a resolution to the fermion doubling problem. We construct and analyze the fermion propagator on the quantum geometry, and we show that all fermion doubler modes are suppressed in the propagator. Our result suggests that the superposition nature of quantum geometry should resolve the tension between fermion and the fundamental discreteness, and relate to the continuum limit of quantum gravity.

\end{abstract}

%\keywords{Black hole, loop quantum gravity, singularity resolution }
\maketitle
%\tableofcontents

The physical world is made of gravity and matters. The quantum theory of matters has been well-developed by the standard model, where gravity is, however, classical. It has been suggested that a consistent theory of gravity coupled to quantized matter should also have the gravitational field quantized \cite{Page:1981aj}. On the other hand, Quantum Gravity (QG) also needs matter couplings, not only because matter fields are dominant in the universe, but also due to the universal gravity-matter interaction, which makes matters serve as probes for empirically testing QG effects. The matter couplings provide a toolbox for early studies of the ultraviolet (UV) behavior of QG \cite{Deser:1974cy,tHooft:1974toh,Goroff:1985th}, and play important roles in cosmology, black holes, asymptotic safety, QG phenomenology, etc \cite{Ashtekar:2021dab,Hawking:1975vcx,Dona:2013qba,Perez:2017krv}.  

Loop Quantum Gravity (LQG) is a promising candidate for the background-independent and non-perturbative QG. Matter couplings in LQG have been extensively explored (see e.g. \cite{PhysRevD.40.2572,Thiemann:1997rt,Sahlmann:2002qj,Lewandowski:2021bkt,Bodendorfer:2011ny,Bojowald:2007nu,Domagala:2010bm,Kisielowski:2018oiv,Bianchi:2010bn,Oriti:2006jk,Mansuroglu:2020acg,Zhang:2011vg,Kami_ski_2006}). A framework of coupling standard model to LQG is developed with the interesting feature of the ultraviolet regularity \cite{Thiemann:1997rt}.

This work focuses on chiral fermions in LQG. Due to the fundamental discreteness of LQG, the fermion coupling resembles the Lattice Field Theory (LFT) to some extent, on any given discrete spacetime in LQG. It is well-known that chiral fermions in LFT suffers the doubling problem, i.e. each fermion results in $2^m$ fermion species on $m$-dimensional lattice \cite{montvay_munster_1994,Nielsen:1981hk}. This fact leads to the suspicion of the fermion doubling problem in LQG \cite{Barnett:2015ara}. However, there also exists the opposite opinion arguing that the fermion doubling problem may not exist in LQG due to the superposition of quantum geometries \cite{Gambini:2015nra,Bianchi:2010bn,Han:2011as}. The confusion on fermion doubling has been long-standing in the LQG community since the first paper on LQG-fermion in 1997. The similar issue should exist in all QG approaches with discrete spacetimes. This issue is crucial because it reveals the tension between fermion and the fundamental discreteness of quantum spacetime, given that the fundamental discreteness is believed to be a key feature of QG \cite{Hooft_2016,1955PhRv...97..511W,Hawking:1978pog}.

A resolution of this tension is proposed in this paper from the perspective of LQG. We construct and compute the chiral fermion propagator of LQG. Due to the superposition of lattices given by the quantum geometry in LQG, the resulting propagator averages the lattice fermion propagators over a sequence of lattice refinements. We show that the average results in suppression of all fermion doubling modes, while the physical mode is kept unchanged. Our result implies that the LQG fermion propagator has the correct continuum limit, and suggests the nature of superposition in quantum geometries be the key to resolve the tension between fermion and the fundamental discreteness. Our result also provides an interesting example for understanding the continuum limit of LQG.

\emph{Fermion doubling and a resolution.}--- Let us firstly consider LFT on 4d Minkowski spacetime, where the time is continuous and the space is discretized by a cubic lattice $\gamma$ \cite{Kogut:1974ag,Creutz:2001wp}. The propagator of chiral fermion has the following expression:
\bee
G_{\omega,\vec{k}}(\gamma,a)=\frac{-\omega\mathbf{1}_{2\times 2}+\frac{1}{a}\sum_{I=1}^{3}\bm{\sigma}^{I}\sin\left(a k_{I}\right)}{\omega^{2}-\frac{1}{a^2}\sum_{I=1}^{3}\sin^{2}\left(a k_{I}\right)+i\epsilon }
\eee
where $I=1,2,3$ labels the directions on $\gamma$. $a$ is the lattice spacing, and $\epsilon$ is the Feynman regulator. $G_{\omega,\vec{k}}(\gamma,a)$ is a $2\times2$ matrix whose matrix indices are Weyl spinor indices, denoted by $A,B=\pm$. The range of momentum $k^I$ is the fundamental Brillouin zone (FBZ) $-\pi/a\leq k^I <\pi/a$. $G_{\omega,\vec{k}}(\gamma,a)$ has the doubling problem: A physical pole at e.g. $(\o,\vec{k})=(\o,k,0,0)$ (satisfying $\omega=\frac{1}{a^2}\sin\left(a k\right)>0$) implies another spurious doubler pole at $(\o,\frac{\pi}{a}-k,0,0)$, so the fermion species is doubled in each direction on the lattice. The fermion doubling problem causes the trouble of the continuum limit of fermions on the lattice, and is intractably linked to chirality by the Nielsen-Ninomiya no-go theorem \cite{Nielsen:1981hk}.

\begin{figure}[ht]
	\begin{center}
	\includegraphics[width=0.4\textwidth]{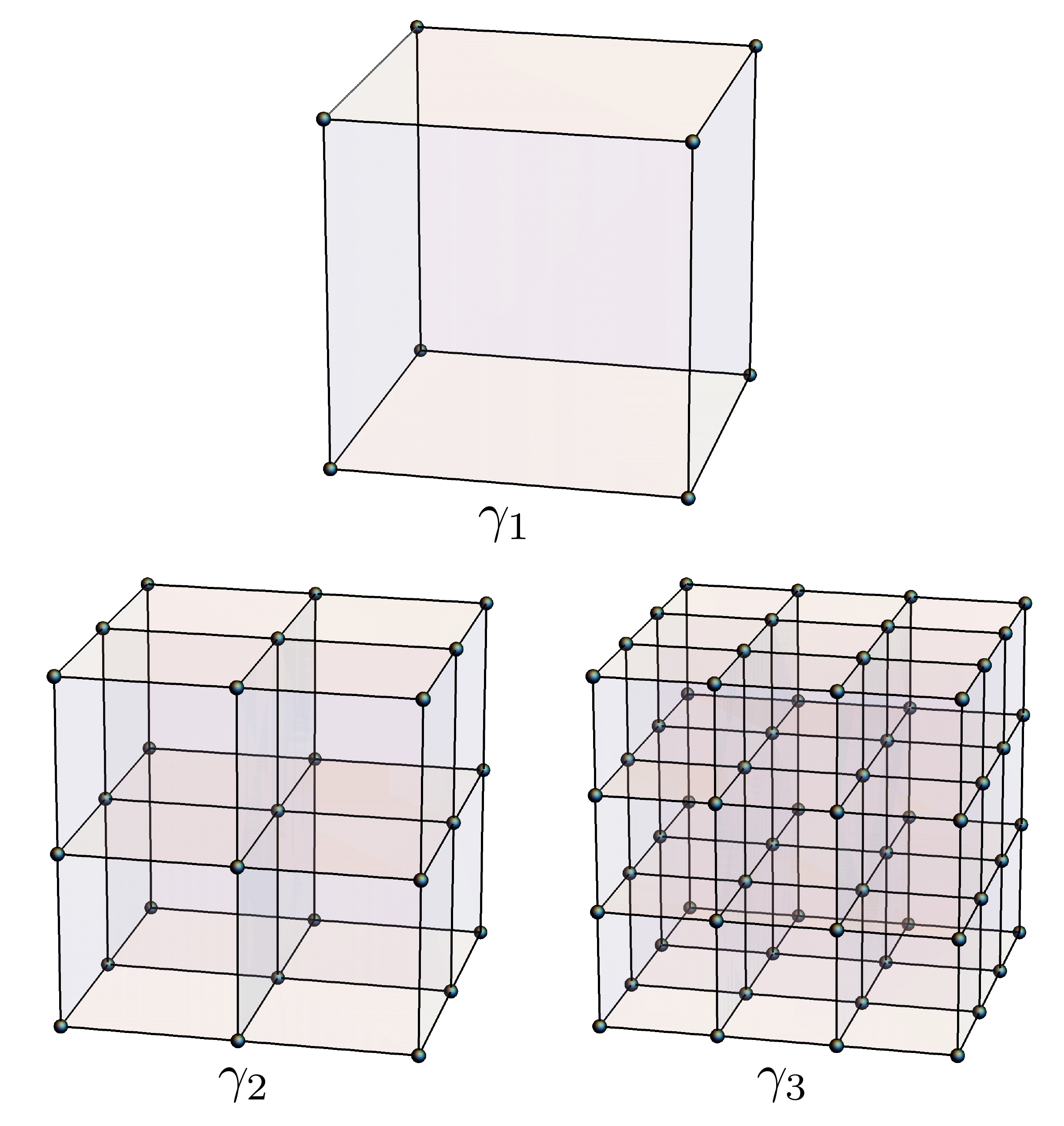}
	\caption{ Lattice refinement of a cell of $\gamma_1$, to $\gamma_2$ and $\gamma_3$. Along each direction on $\gamma_n$, the number of vertices $N_n$ satisfies $N_n=n N_1$.
	}
	\label{refinement}
	\end{center}
\end{figure}

LQG gives quantum geometry states as superposition of lattices, and it indicates a resolution to the doubling problem by taking into account the average of the propagators over the refinements of the lattice. Namely we consider the following quantity, whose derivation from LQG is given in a moment
\bee
{G}_{\o,\vec{k}}=\frac{1}{L}\sum_{n=1}^L\chi_{n,\vec{k}}G_{\omega,\vec{k}}\lt(\gamma_n,a_n\rt),\label{average00}
\eee
where $L$ is large and $\ga_n$ are refinements of $\ga_1=\ga$ and has the spacing $a_n=a_1/n$ (see FIG.\ref{refinement}). We introduce the short-hand notation $\mathrm{FBZ}(\ga_n)\equiv[-\pi/a_n,\pi/a_n)^3$ for the FBZ on $\gamma_n$ with the spacing $a_n$, and we have $\mathrm{FBZ}(\ga_n)\subset \mathrm{FBZ}(\ga_{n+1})$. $\chi_{n,\vec{k}}=1$ if $\vec k$ belongs to $\mathrm{FBZ}(\ga_n)$, otherwise $\chi_{n,k}=0$.

\begin{figure}[ht]
	\begin{center}
	\includegraphics[width=0.5\textwidth]{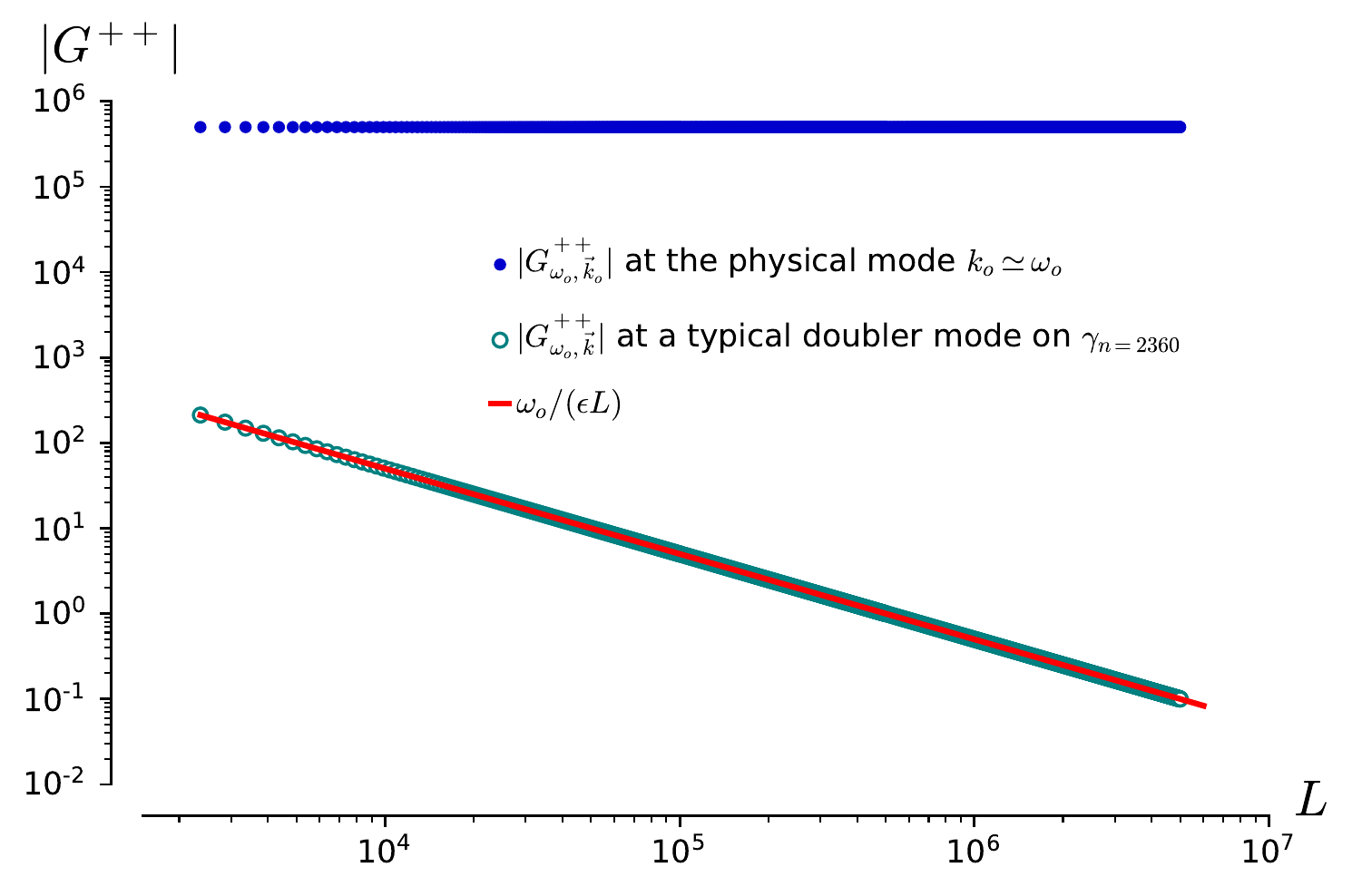}%{poles1.pdf}
	\caption{Log-log plots of  $|G^{++}_{\o_o,\vec{k}_o}|$ at the physical mode $k_o\simeq \o_o$ (blue dots), and $|G^{++}_{\o_o,\vec{k}}|$ at a typical doubler mode on $\ga_{n}$ with $n=2360$ (green circles). The parameters are $\o_o=50,a_1=2^{-50}$ $\epsilon=10^{-4}$. The red line draws $\o_o/(\epsilon L)$ as a function of $L$
	}
	\label{poles}
	\end{center}
\end{figure}

The assumptions of Nielsen-Ninomiya theorem is clearly violated by summing over lattices. Although apparently \eqref{average00} adds to ${G}_{\o,\vec{k}}$ the spurious doubler poles from all $\ga_n$, the contribution from each doubler pole are suppressed, thanks to the refinement and averaging. To explain the mechanism, we firstly note that for $\epsilon> 0$, the poles of ${G}_{\o,\vec{k}}$ are away from real $(\o,\vec{k})$-space, so ${G}_{\o,\vec{k}}$ is finite for real $(\o,\vec{k})$'s. Given an $\o_o>0$ such that the physical mode, e.g.  {$\vec{k}_o=(k_o,0,0)$, is in $\mathrm{FBZ}(\ga_1)$}, and assuming $a_1$ is sufficiently small so that $k_o\simeq \o_o$, $(\o_o,\vec{k}_o)$ is close to a pole of every $G_{\omega,\vec{k}}\lt(\gamma_n,a_n\rt)$ and gives equally large contributions to all terms in \eqref{average00}, while the contributions are averaged over $\{\ga_n\}$. In contrast, each doubler mode $\vec{k}_n=({\pi}/{a_n}-k_o,0,0)$ associated to $\ga_n$ only gives a single large contribution to the $n$-th term in \eqref{average00}, because different terms in \eqref{average00} has different doubler poles and $\vec{k}_n$ is only close to one of them. Then $G_{\omega_o,\vec{k}_n}$ only receives the dominant contribution from one term, and thus it is suppressed by the overall $1/L$. The numerical experiment with $\omega_o=50,a_1=2^{-50},L=1.2\times10^5,\epsilon=10^{-4}$ shows that $|G^{++}_{\omega_o,\vec k}|<4.2$ at the doubler modes on all $\{\ga_n\}_{n=1}^L$, while $|G^{++}_{\omega_o,\vec k_o}|\simeq 5\times 10^5$ (equals $\o_o/\epsilon$) at the physical mode. FIG.\ref{poles} demonstrates that as $L$ goes large, $G_{\omega_o,\vec{k}}$ remains large and constant at the physical mode $k_o$, while $G_{\omega_o,\vec{k}}$ is suppressed at the doubler mode. The coincidence between $G_{\omega_o,\vec{k}_n}$ and $\omega_o/(\epsilon L)$ shown in FIG.\ref{poles} indicates that $G_{\omega_o,\vec{k}}$ at the doubler mode indeed only receives the dominant contribution from one term in the sum.

In absence of doubling mode, $G_{\o_o,\vec{k}}$ is peaked at the physical mode $\vec{k}=\vec{k}_o$. In any neighborhood of $\vec{k}_o$ and with sufficiently large $L$, $G_{\o_o,\vec{k}}$ approximates the continuum fermion propagator arbitrarily well, due to the well-known result $\lim_{L\to\infty}\frac{1}{L}\sum_{n=1}^L f_n=\lim_{n\to\infty}f_n$ for any sequence $\{f_n\}$.

We can generalize \eqref{average00} and consider
\bee
G_{\omega,\vec{k}}^{(w)}=\sum_{n=1}^L|w_n|^2\chi_{n,\vec{k}}G_{\omega,\vec{k}}(\ga_n,a_n),\label{average01}
\eee
with the generic weight $|w_n|^2$ satisfying $\sum_{n=1}^L|w_n|^2=1$. $G_{\omega,\vec{k}}^{(w)}$ reduces to $G_{\omega,\vec{k}}$ when the constant weight $|w_n|^2=1/L$ is chosen. The mechanism of suppressing the doubler modes does not rely on the specific choice of $w_n$. Thus our result also applies to $G_{\omega,\vec{k}}^{(w)}$ with generic $w_n$.

\emph{LQG coupled to chiral fermion.}--- In the following, we show that the propagator \eqref{average00} can be derived from LQG with fermion coupling. The 4d spacetime topology is assumed to be $\mathbb{R}\times\Sigma$. A graph $\ga$ in the spatial slice $\Sig$ consists of oriented edges $e$ and vertices $v$ as sources and targets of the edges. The LQG Hilbert space $\ch$ is given by the direct sum over all graphs $\ch=\oplus_\ga\ch_\ga/\mathcal{G}$, where $\ch_\ga$ is given by $\ch_{\ga}=\ch_\ga^G\otimes \ch_\ga^F$, where $\ch_\ga^G$ is spanned by spin-networks on $\ga$ with nonzero spins, and $\ch_\ga^F=\otimes_{v}\ch_v^F$ ($\ch_v^F\simeq \mathbb{C}^4$) is the Hilbert space of fermions. $\oplus_\ga\ch_\ga$ carries the representation of the holonomy-flux algebra \cite{lewandowski2005uniqueness,ashtekar1994representation} $[\hat{h}_e,\hat{h}_{e'}]=0$, $[\hat{h}_e,\hat{p}^a_{e'}]=\b\ell_P^2\delta_{e,e'}(\bm{\sig}^a/2)\hat{h}_e$, { $[\hat{p}^a_e,\hat{p}^b_{e'}]=i\b\ell_P^2\delta_{e,e'}\eps^{abc}\hat{p}^c_e$} ($\ell_P^2=8\pi G\hbar$ and $\beta$ is the Barbero-Immirzi parameter) and the Weyl fermion oscillators $[\hat{\zeta}_{v}, \hat{\zeta}_{v'}^{\dagger}]_{+}=\delta_{v, v^{\prime}}\mathbf{1}_{2\times 2} $. $\cg$ are SU(2) gauge transformations, and $\ch$ only contains SU(2) gauge-invariant states. We assume the spatial slice $\Sig\simeq\mathbb{T}^3$ and the periodic boundary condition.

There is a physical Hamiltonian acting on $\ch$ \cite{Thiemann:2020cuq,Giesel:2012rb,giesel2010algebraicIV} 
\bee
\hat{\bf H}=\oplus_\ga \hat{\bf H}_\ga,\quad \hat{\bf H}_\ga=\hat{\bf P}_\ga(\hat{H}^G_\ga+\hat{H}^F_\ga)\hat{\bf P}_\ga, \label{Hoplus}
\eee 
where each $\hat{\bf P}_\ga$ is the projection onto $\ch_\ga$. $\hat{H}^G_\ga$ is identical to the graph-preserving LQG Hamiltonian constraint operator with unit lapse \cite{Giesel:2006uj,giesel2007algebraic}. $\hat{H}^F_\ga$ is the graph-preserving fermion Hamiltonian \cite{Lewandowski:2021bkt,Thiemann:1997rt,Sahlmann:2002qj,Giesel:2006uj}. $\hat{H}^F_\ga$ is quadratic in $\hat{\zeta}_v$: $\hat{H}^F_\ga/\hbar=\sum_{\lag v,v'\rag}\hat{\zeta}_v^\dagger M_{v,v'}(\hat{h},\hat{p})\hat{\zeta}_{v'}$ where $M_{v,v'}$ is a $2\times 2$ matrix with entries involving only quantum-geometry operators. The detailed expression of $\hat{H}^G_\ga+\hat{H}^F_\ga$ is reviewed in {Appendix} A. The theory quantizes Einstein gravity coupling to the Weyl fermion and Gaussian dust in the reduced phase space formulation (relational framework) \cite{Kuchar:1990vy,Giesel:2012rb,Han:2020chr,Dittrich:2005kc,Thiemann:2004wk,Rovelli:2001bz,Rovelli:1990ph}. Semiclassically $h_e,p_e^a$ are Dirac observables being the holonomy and flux of the Ashtekar-Barbero variables in the reference frame defined by the Gaussian dust. As a promising aspect, the theory is free of the complications from the quantum Hamiltonian and diffeomorphism constraints, because it quantizes the reduced phase space, where both constraints are resolved at the classical level. All quantities in the theory are Dirac observables from the start.

\emph{Initial state.}--- Based on the above framework of fermion coupling in LQG, the fermion propagator are going to be constructed, and shown to recovers \eqref{average01} in the semiclassical approximation. As an important ingredient in the fermion propagator, $|\O\rangle\in\ch$ coupling quantum geometry and fermions needs to be constructed as the LQG analog of the fermion ground state on the semiclassical flat spacetime. Here we define $|\O\rangle$ to be an entangled state being the superposition of $|\Omega'_{\gamma}\rangle\equiv\hat{\mathbf{P}}_\ga|\psi_{\ga}\rangle\otimes|\o_\ga\rangle\in \ch_\ga^G\otimes\ch_\ga^F$ over many different graphs $\ga$, modulo SU(2) gauge transformations. On each $\ga$, the quantum geometry state $|\psi_{\ga}\rangle$ is Thiemann's coherent state peaked at the flat spacetime geometry \cite{thiemann2001gaugeII,thiemann2001gaugeIII}. $|\psi_{\ga}\rangle$ endows $\ga$ with the semiclassical flat geometry and vanishing extrinsic curvature. $|\o_\ga\rangle$ is the normalized fermion ground state on the semiclassical flat spacetime, and associates to the lowest energy level of the \emph{effective} fermion Hamiltonian $\hat{ H}_{\ga}^F(\psi_\ga)/\hbar=\|\psi_{\ga}\|^{-2}\sum_{\lag v,v'\rag}\hat{\zeta}_v^\dagger\langle\psi_{\ga}| M_{v,v'}(\hat{h},\hat{p})|\psi_{\ga}\rangle\hat{\zeta}_{v'}$. $\hat{ H}_{\ga}^F(\psi_\ga)$ equals the usual fermion Hamiltonian on the flat lattice plus $O(\ell_P^2)$ corrections (see Appendix B for more details, see also \cite{Sahlmann:2002qj} for the early study of the similar idea). 

Any smooth geometry admits many different discretizations based on different graphs. We propose that the semiclassical state of the smooth geometry should be a superposition of states on different graphs, where each state is semiclassical on one graph and relate to the discretization of the smooth geometry on the graph. Here, we consider the smooth geometry to be flat and choose cubic graphs for the satisfactory semiclassical properties at the discrete level \cite{giesel2007algebraic,Flori:2008nw}. We consider a set of cubic graphs $\{\ga_n\}_{n=1}^L$, where $\ga_{n}$ is a refinement of $\ga_1$ by subdividing every cube into $n^3$ cubes. $N_n=nN_1$ is the number of vertices on $\ga_n$ in every direction. $N_1$ is assumed to be even. We define $|\O'\rangle=
\sum_{n=1}^Lw_n|\O'_{\gamma_n}\rangle$, where $|\O'_{\gamma_n}\rangle=\hat{\mathbf P}_{\ga_n}|\psi_{\ga_n}\rangle\otimes |\o_{\gamma_n}\rangle$, and the gauge invariant projection
\bee
|\O\rangle=
\sum_{n=1}^Lw_n|\O_{\ga_n}\rangle,
\ \  |\O_{\ga_n}\rangle=\hat{\bf P}_{\cg}%{\color{red}{|\Omega_\gamma'\rangle}}
\lt( \hat{\mathbf P}_{\ga_n}|\psi_{\ga_n}\rangle\otimes |\o_{\gamma_n}\rangle\rt).\nonumber
\eee
where $\hat{\bf P}_{\cg}$ is the (group-averaging) projection onto $\ch$. $|\O_{\ga_n}\rangle$ are mutually orthogonal and $\|\O_{\ga_n}\|=1$. $L$ is finite and large. The weight $w_n$ satisfies $\sum_{n=1}^L|w_n|^2=1$ so that $\|\O\|=1$. The discrete geometries on $\{\ga_n\}_{n=1}^L$ given by $\{|\psi_{\ga_n}\rangle\}_{n=1}^L$ are discretizations of the smooth flat geometry, and give the lattice spacings $\{a_n\}_{n=1}^L$ satisfying $a_n=a_1/n$. Indeed, the discrete geometry on $\gamma_n$ are implied by the expectation values $\langle\psi_{\ga_n}|\hat{h}_{e}|\psi_{\ga_n}\rangle=\mathbf{1}_{2\times 2}+O(\ell_P^2)$ and $\langle\psi_{\ga_n}|\hat{p}^a_{e(I)}|\psi_{\ga_n}\rangle=a^2_n\delta^a_I+O(\ell_P^2)$, where $e(I)$ is the edge along the $I$-direction. The total spatial volume $V_{\rm tot}=a_nN_n$ is $n$-independent. %More details about $|\O\rangle$ are given in Appendix B.

\emph{Fermion propagator.}--- We need the local field operator $\hat{\Psi}(v)$ in order to construct the fermion propagator. However, classically the fermion field $\Psi$ and $\zeta$ are not the same but related by $\zeta=\sqrt[4]{\det(q)}\Psi$, where $\sqrt{\det(q)}$ is the spatial volume density \cite{Thiemann:1997rq,Thiemann:1997rt}. Motivated by the classical relation, we define the fermion field operator by $\hat{\Psi}(v)=\hat{\zeta}_v\hat{V}_v^{-\frac{1}{2}}$, where $\hat{V}_v^{-\frac{1}{2}}$ is the square-root inverse volume (see {Appendix} A or \cite{Giesel:2005bk,yang2016new}). We define the Heisenberg operator $\hat{\Psi}(\t,v):=e^{\frac{i}{\hbar}\hat{\bf H}(\t-T_i)}\hat{\Psi}(v) e^{-\frac{i}{\hbar}\hat{\bf H}(\t-T_i)}$ and similarly for $\hat{\Psi}^\dagger(\t,v)$, where $T_i$ is the initial time. We prepare the initial state $|\O\rangle$ at $T_i$ and define the fermion propagator by 
\bee
\mathscr{G}^{AB}(\t_1,\t_2;v_1,v_2)=\langle\O'|\ct[\hat{\Psi}^A(\t_1,v_1)\hat{\Psi}^B{}^\dagger(\t_2,v_2)]|\O\rangle,\nonumber
\eee
where $\ct$ is the time-ordering. Here $v_1,v_2$ are vertices in $\ga_1$, and thus they belong to all the refinements $\ga_n$. $\mathscr{G}^{AB}$ is not SU(2) gauge invariant, as the usual situation in gauge theories. An example of the gauge invariant observable is 
{$\mathscr{G}_h(\t_1,\t_2;v_1,v_2)=\langle\O|(\hat{h}_{c})_{BA}\ct[\hat{\Psi}^A(\t_1,v_1)\hat{\Psi}^B{}^\dagger(\t_2,v_2)]|\O\rangle$}, where $\hat{h}_{c}$ is the holonomy operator along a path $c\subset \ga_1$ connecting $v_1,v_2$ \footnote{In $\mathscr{G}_h$, \unexpanded{$\langle\O|$} can be replaced by \unexpanded{$\langle\O'|$} for free, because the operator is gauge invariant and $\p_{\cg}^2=\p_\cg$.}. In the following, we proceed to compute $\mathscr{G}^{AB}$, then we show $\mathscr{G}_h\simeq\mathscr{G}^{AB}\delta_{AB}$ up to $O(\ell_P^2)$.

Since $\hat{\bf H}$ is a direct sum and $|\O_{\ga_n}\rangle$ are mutually orthonormal, $\mathscr{G}^{AB}$ is a sum of the fermion propagators on graphs $\{\ga_n\}$
\bee
\mathscr{G}^{AB}(\t_1,\t_2;v_1,v_2)&=&%\frac{1}{L}
\sum_{n=1}^L|w_n|^2\mathscr{G}_n^{AB}(\t_1,\t_2;v_1,v_2)\label{GAB0}\\
\mathscr{G}_n^{AB}(\t_1,\t_2;v_1,v_2)&=&\langle\O'_{\ga_n}|\ct[\hat{\Psi}^A(\t_1,v_1)\hat{\Psi}^B{}^\dagger(\t_2,v_2)]|\O_{\ga_n}\rangle\nonumber
\eee
%$\mathscr{G}_n^{AB}$ is the fermion propagator on single graph $\ga_n$. 
We compute $\mathscr{G}_n^{AB}$ with the time-order $\t_1>\t_2$, while the computation for $\t_1<\t_2$ is similar. We apply the coherent-state path integral method by discretizing the time-evolution operator $e^{-\frac{i}{\hbar}\hat{\bf H}_\ga(\t-T_i)}\simeq[1{-\frac{i}{\hbar}\hat{\bf H}_\ga\delta\t}]^{j_{\t}}$ with $j_\tau\delta\t={\t-T_i}$ and large $j_\tau$, and inserting at each step the over-completeness relation of the coherent state $\int \rmd Z_j(\ga_n)|Z_j(\ga_n)\rangle\langle {Z_j(\ga_n)}|=1$ ($j=0,\cdots,j_\t$) with $|Z(\ga_n)\rangle=|\psi_{(h,p)(\ga_n)}\rangle\otimes|\phi_{\zeta({\ga_n})}\rangle$, where $|\psi_{(h,p)(\ga_n)}\rangle$ is Thiemann's coherent state of quantum geometry, and $|\phi_{\zeta({\ga_n})}\rangle$ satisfying $\hat{\zeta}_v|\phi_{\zeta({\ga_n})}\rangle=\zeta_v|\phi_{\zeta({\ga_n})}\rangle$ is the standard coherent state of the fermion oscillators. $Z(\gamma_n)$ is a short-hand notation of the data $\{h_{e},p_{e}^a,\zeta_{v}\}_{e,v}$ where the coherent state is peaked. $\rmd Z(\ga_n)=\rmd g(\gamma_n)\prod_{v,A}\rmd \zeta^A_{v}{}^*\rmd\zeta^A_{v}$ and $\rmd g(\gamma_n)$ denotes the measure of $\{h_e,p_e^a\}_e$ \cite{thiemann2001gaugeII}. Following the standard derivation, we obtain the path integral: 
\bee
\mathscr{G}_n^{AB}&&=\int\prod_{j=0}^{N}\rmd Z_j(\ga_n)\, e^{S^{G}_{\ga_n}(h,p)+S_{\ga_n}^{F}(h,p,\zeta,\zeta^*)}\label{pathintegral00}\\
&&(V^{-\frac{1}{2}}_{v_1}{\zeta}_{v_1}^A)_{j_1}(V^{-\frac{1}{2}}_{v_2}{\zeta}^{B*}_{v_2})_{j_2}\langle\O'_{\ga_n}|Z_{N}(\ga_n)\rangle\langle Z_0(\ga_n)|\O_{\ga_n}\rangle\nonumber
\eee
where $j_{1,2}=j_{\tau_{1,2}}=(\t_{1,2}-T_i)/\delta\t$ and $N=2j_{1}$. $\langle\O'_{\ga_n}|Z_{N}(\ga_n)\rangle$ and $\langle Z_0(\ga_n)|\O_{\ga_n}\rangle$ are linear in $\zeta_v$ and $\zeta_v^*$ respectively , and the fermion action $S^F_{\ga_n}$ is quadratic in $\zeta_v$
\bee
S_{\ga_n}^F&=&\sum_{j,j',v,v'}\zeta^\dagger_{v,j}\mathscr{H}_{v,j;v',j'}(h,p)\zeta_{v',j'}\nonumber\\
\mathscr{H}_{v,j;v',j'}&=&\delta_{v',v}\delta_{j,j'+1}-\delta_{j',j}\delta_{v',v}\nonumber\\
&&-i\rho_j\delta_{j,j'+1}{\delta\tau}\lt\langle\psi_{g_j(\ga_n)}\lt| M_{v,v'}(\hat{h},\hat{p})\rt|\psi_{g_{j'}(\ga_n)}\rt\rangle,\nonumber
\eee
where $\rho_j=1$ or $-1$ at instances before or after $\t_1$. We perform the Gaussian integral of the fermion and obtain the following expression:
\bee
\!\!\!\!\mathscr{G}_n^{AB}=\int\limits_{\mathcal{G}}\rmd u\int\prod_{j=0}^{N}\rmd g_j(\ga_n)\,e^{S^{G}_{\ga_n}}(V^{-\frac{1}{2}}_{v_1})_{j_1}(V^{-\frac{1}{2}}_{v_2})_{j_2}\,\ci^{AB}_{n}\label{gintegral}
\eee
where $\ci^{AB}_{n}=\ci^{AB}_n(p,h,u)$, resulting from the fermion Gaussian integral, is independent of $\hbar$. The integral over SU(2) gauge transformations $u$ comes from $\mathbf{P}_{\cg}$ in $|\O_{\ga_n}\rangle$. %For the integral is over gravity variables $h_{e,j},p_{e,j}^a$, 
The effective action of geometry $S^{G}_{\ga_n}$ scales as $1/\ell_P^2$ as $\ell_P\to 0$ \cite{han2020effective} (see also Appendix C). So the stationary phase approximation can be employed to study the semiclassical approximation of \eqref{gintegral}. It is shown in \cite{han2020effective,Han:2020chr} that the equations of motion (EOMs) $\delta S^{G}_{\ga_n}=0$ reproduce the classical Hamilton's equation of holonomy and fluxes ${\rmd h_e}/{\rmd \tau}=\{h_e,H^G\}$,  ${\rmd p^a_e}/{\rmd \tau}=\{p^a_e,H^G\}$, where $H^G$ is the discrete gravity Hamiltonian on $\ga_n$ with unit lapse and vanishing shift. The only solution $(\mathring{h},\mathring{p},\mathring{u})$ satisfying the EOMs and compatible to the initial state $|\O_{\ga_n}\rangle$ is $\mathring{u}=\mathbf{1}$ and the flat spacetime geometry, where $\mathring{h}_{e,j}=\mathbf{1}_{2\times 2}$ and $\mathring{p}_{e(I),j}^a=a_n^2\delta^a_I$, i.e. the lattice geometry is constantly flat with spacing $a_n$ at all time. Eq.\eqref{gintegral} has the following semiclassical approximation
\bee
\mathscr{G}_n^{AB}&=&a_n^{-3}\ci_n^{AB}(\mathring{p},\mathring{h},\mathring{u})\int\limits_{\mathcal{G}}\rmd u\int\prod_{j=0}^{N}\!\rmd g_j(\ga_n)\, e^{S^{G}_{\ga_n}}\lt[1+O(\ell_P^2)\rt]\nonumber\\
&=&\,a_n^{-3}\ci_n^{AB}(\mathring{p},\mathring{h},\mathring{u})\lt[1+O(\ell_P^2)\rt].\label{IAB}
\eee
Here the integral removes all fermions degrees of freedom in $\mathscr{G}_n$, so it equals $\langle\tilde{\psi}_{\gamma_n}|\tilde{\psi}_{\gamma_n}\rangle$, where $|\tilde{\psi}_{\gamma_n}\rangle\equiv\mathbf{P}_\cg\mathbf{P}_{\gamma_n}|\psi_{\gamma_n}\rangle$. $\langle\tilde{\psi}_{\gamma_n}|\tilde{\psi}_{\gamma_n}\rangle=1+O(\ell_P^2)$ is implied by $\|\O_{\gamma_n}\|=1$ and $\|\o_{\gamma_n}\|=1$. More details about the semiclassical analysis of $\mathscr{G}_n^{AB}$ can be found in {Appendix} C. 

The gauge invariant observable $\mathscr{G}_h$ inserts an $({h}_c)_{AB}$ in the above path integral. Then the stationary phase approximation is \eqref{IAB} contracted by $(\mathring{h}_c)_{AB}=\delta_{AB}$.

The geometry is freezed to be flat in \eqref{IAB}, and $I_n^{AB}$ reduces to the standard fermion propagator on the flat lattice with spacing $a_n$:
\bee
\mathscr{G}_n&=&\frac{1}{V_{\rm tot}}\sum_{\vec{k}\in \text{FBZ}(\ga_n) }\int_{\mathbb{R}}\frac{\rmd\omega}{2\pi i}\, e^{-i\omega(\tau_1-\t_2)+ia_n \vec{k}\cdot\left(\vec{v}_{1}-\vec{v}_{2}\right)}\nonumber\\
&&G_{\omega,\vec{k}}(\ga_n,a_n)\lt[1+O(\ell_P^2)\rt],\label{IAB1}
%&&\frac{\omega\mathbf{1}_{2\times 2}-\frac{1}{a_n}\sum_{I=1}^{3}\bm{\sigma}^{I}\sin\left(a_nk_{I}\right)}{\omega^{2}-\frac{1}{a_n^2}\sum_{I=1}^{3}\sin^{2}\left(a_n k_{I}\right)+i\epsilon }\lt[1+O(\ell_P^2)\rt],\label{IAB1}
\eee
where the continuous limit of time has been taken. The lattice vertices $\vec{v}_i\in\mathbb{Z}^3$ and $\vec{v}_i=\vec{v}_i+N_n$ ($i=1,2$). The lattice Fourier mode ${k}^I\in2\pi \mathbb{Z}/(N_1a_1)$ is summed over the FBZ on $\ga_n$ with the spacing $a_n$. %$V_{\rm tot}=(\sqrt{p_n}N_{n})^{3}$ is the total spatial volume of the flat geometry and is constant for all $\ga_n$. $\epsilon>0$ is the Feynman regulator. 

Insert \eqref{IAB1} in \eqref{GAB0}, and extend the sum of $\vec{k}$ to FBZ on the finest lattice ($\text{FBZ}(\gamma_n)\subset \text{FBZ}(\gamma_L)$ for all $n<L$). Notice that $\vec{x}_{1}-\vec{x}_{2}\equiv a_n(\vec{v}_{1}-\vec{v}_{2})$ is $n$-independent. The LQG fermion propagator $\mathscr{G}$ becomes
\bee
\mathscr{G}&=&\frac{1}{V_{\rm tot}}\sum_{k^{I}\in\text{FBZ}(\gamma_L)}\int_{\mathbb{R}}\frac{\rmd\omega}{2\pi i}\, e^{-i\omega(\tau_1-\t_2)+i\vec{k}\cdot\left(\vec{x}_{1}-\vec{x}_{2}\right)}\mathscr{G}_{\o,\vec{k}},\nonumber\\
\mathscr{G}_{\o,\vec{k}}&=&\sum_{n=1}^L|{w}_n|^2\chi_{n,\vec{k}}G_{\omega,\vec{k}}(\ga_n,a_n)\lt[1+O(\ell_P^2)\rt].\label{scrGok}
\eee
Up to $O(\ell_P^2)$ corrections, $\mathscr{G}_{\o,k}$ recovers \eqref{average01} in which the fermion doublers are suppressed for large $L$.

\emph{Discussion.}--- Our analysis shows that the quantum geometry as the superposition of states with different lattices results in that the fermion propagator from LQG averages the fermion propagators on the lattices. In the resulting fermion propagator, all doubler modes on the lattices are suppressed by the average, while the physical mode are kept unchanged. Our result indicates that the fermion in LQG is free of the doubling problem. It also suggests that the superposition nature of quantum geometry should be the key to resolve the tension between fermion and the fundamental discreteness in QG. 

The fact that the superposition of lattices brings the fermion propagator close to the continuum limit suggests that the continuum limit of the full LQG should also relate to the superposition of lattices. This is similar to the approach of group field theory \cite{Finocchiaro:2020fhl}.

Interestingly, the propagator $\mathscr{G}_{\o,\vec{k}}$ suggests that the quantum geometry provide a soft UV cut-off to fermions. %Indeed, our discussion of $G_{\omega,\vec{k}}$ has focused on $\o_o$ whose physical mode $\vec{k}_o$ is in $ \mathrm{FBZ}(\ga_1)$. 
Let us scale $\vec{k}$ large such that $\vec{k}$ is outside $\mathrm{FBZ}(\ga_m)$ for certain $m$, then all terms with $n\leq m$ in \eqref{scrGok} vanish because of $\chi_{n,\vec k}$. Hence, as we scale $\vec{k}$ large, $|\mathscr{G}^{AB}_{\o,\vec{k}}|$ is suppressed because less and less terms survive in the sum. This result is consistent with the expectation that QG should regularize the UV behavior of matter fields.

We have set the upper bound $L$ of the sum to be finite, so that the lattice refinement gives maximally $(LN_1)^3$ vertices. $L$ actually correlates to $\ell_P$, and $L$ has to be finite as far as $\ell_P$ is finite. The technical reason is the following: The $O(\ell_P^2)$ correction in $\mathscr{G}_n$ may become non-negligible, when the spacing becomes comparable to the Planck length, i.e. $a_n\sim \ell_P$ \cite{giesel2007algebraic,Zhang:2021qul}. Fixing $V_{\rm tot}$, $L$ has to be bounded by $L_0= V_{\rm tot}^{1/3}/(N_1\ell_P)$ in order that the semiclassical approximation can be applied to all $\mathscr{G}_n$ in the sum. Therefore, the refinement limit $L\to\infty$ can only be taken together with the semiclassical limit $\ell_P\to 0$. Further investigation is needed to go beyond the semiclassical approximation when computing $\mathscr{G}_n$. Then one may consider $L\to\infty$ without the semiclassical limit.

%Recall FIG.\ref{poles}, when we choose $L\sim L_0$, the contribution of the doubler mode to the propagator is approximately $\o_o/(\epsilon L) \sim \ell_P\o_oN_1/(\epsilon V_{\rm tot}^{1/3})$, negligible in the semiclassical approximation. 

Our result shares the similarity with the random LFT \cite{Christ:1982zq,Pang:1986kj,Perantonis:1987fs,Griffin:1992ni} by the average over lattices. But a key difference is that we sum over lattices with different numbers of vertices relating to the lattice refinement, while the random lattice often fixes the number of vertices. It is interesting to further explore the relation to LFT, in order to see the impact from quantum geometry to high energy particle physics. Our computation should extend to n-point functions and further probe the UV behavior. It is also important to study the chiral anomaly on quantum geometry, which is a research undergoing.

\begin{acknowledgements}

%\emph{Acknowledgements.}--- 

M.H. acknowledges Chen-Hung Hsiao for discussions at early stage of this work. M.H. receives support from the National Science Foundation through grants PHY-1912278 and PHY-2207763. M.H. also acknowledges funding provided by the Alexander von Humboldt Foundation for his visit at the Friedrich-Alexander-Universit\"at Erlangen-N\"urnberg. C. Z. acknowledges the support by the Polish Narodowe Centrum Nauki, Grant No. 2018/30/Q/ST2/00811, and NSFC with Grants No. 11961131013 and No. 11875006.

\end{acknowledgements}

\onecolumngrid

\appendix
\section{A. Hamiltonian opertors }\label{app:HamiltonianAndVolume}

In LQG, to promote the classical Hamiltonian of gravity $H^G\equiv E-(1+\beta^2)L$ to the operator, we need to regularize $H^G$ based on a graph $\gamma$, and express $H^G$ in terms of the basic variables $h_{e}$ and $p_e^a$. %Thiemann's trick \cite{Thiemann:1997rq} is used to obtain the regularized $H^G$ involving the Poisson bracket between $h_e$ and the volume $V_v$ associated to $v$. 
Consider $\gamma$ to be a cubic lattice, and denote by $e(I)$ the edge along the $I$-direction. The quantization of $H^G\equiv E-(1+\beta^2)L$ gives the graph-preserving operator $\hat{H}^G_\gamma\equiv \hat{E}_\gamma-(1+\beta^2)\hat{L}_\gamma$, where
%We use the Thiemann's trick \cite{}, and rewrite, saying, the Euclidean part $E$ in terms of $h_e$, volume $V_v$ of a region dual to vertices $v$ and Poisson bracket. Then the operator $\hat E$ is obtained by replacing the classical variables by their operator correspondences and the Poisson bracket by $[\cdot,\cdot]/(i\hbar)$. Finally the result of $\hat E_\gamma$ based on $\gamma$ reads 
\begin{equation}
\begin{aligned}
\hat E_\gamma&=\frac{-1}{4i\kappa\beta\ell_P^2}\sum_{v\in V(\gamma)}\sum_{e(I),e(J),e(K) \text{ at } v}\epsilon^{IJK}\tr(h_{\alpha_{IJ}}h_{e(K)}^{-1}[h_{e(K)},\hat V_v]).\\
\hat L_\gamma&=-\frac{4\kappa}{i \beta ^7 \ell_P^{10}}\sum_{v\in V(\gamma)}\sum_{e(I),e(J),e(K) \text{ at } v}\varepsilon^{IJK}\tr( [h_{e(I)},[\hat V_\gamma,\hat E_\gamma]]h_{e(I)}^{-1} [h_{e(J)},[\hat V_\gamma,\hat E_\gamma]]h_{e_J}^{-1}[h_{e(K)},\hat V_v]h_{e(K)}^{-1})
\end{aligned}
\end{equation}
$\alpha_{IJ}$ is a minimal loop in $\gamma$ containing edges $e(I)$ and $e(J)$. $\hat V_v$ is the volume operator associated to the vertex $v$, and $\hat V_\gamma=\sum_{v\in V(\gamma)}\hat V_v$ is the total volume operator. 

The fermion Hamiltonian operator is given by $
\hat  H_\gamma^F=\sum_{\langle v,v'\rangle}\hat\zeta_v^\dagger M_{v,v'}(\hat h,\hat p)\hat\zeta_{v'}$, where $ M_{v,v'}(\hat h,\hat p)$ reads
\begin{equation}\label{eq:HF}
\begin{aligned}
 &M_{v,v'}(\hat h,\hat p)=\delta_{v,v'}\hat V_v^{-\frac12}\left(-\frac{(1+3\beta^2)}{4\beta}\sum_{e \text{ at } v} \hat p_e^{a}{\bm \sigma}_a-\frac{\kappa}{i\ell_P^2\beta}\left[\hat E_\ga,\hat V_v\right]\mathbf{1}_{2\times 2}
 \right)\hat V_v^{-\frac12}\\
 &-\frac{\beta-i}{4}\left(\sum_{e\text{ at } v}\delta_{v',t_e} \hat V_v^{-\frac12}{\bm \sigma}_a \hat h_e\hat p_{e}^a\hat V_v^{-\frac12}\right)
 -\frac{\beta+i}{4}\left(\sum_{e' \text{ at } v'}\delta_{v,t_{e'}}\hat V_v^{-\frac12}\hat p_{e'}^ah_{e'}^{-1} {\bm \sigma}_a \hat V_v^{-\frac12}\right)
 \end{aligned}
\end{equation}
where $t_e$ denotes the target of the edge $e$. 
Here $\hat V_v^{-\frac12}$, densely defined on $\ch$, is not the inverse of $\sqrt{\hat V_v}$. Indeed, $\sqrt{\hat V_v}$ takes $0$ as its eigenvalue, so its inverse is not densely defined. $\hat V_v^{-\frac12}$ is obtained by regularizing $\det(e)/\det(q)$ in a neighorhood at $v$ ($\det(e)$ is the determinant of cotriad) \cite{Giesel:2005bk,yang2016new}. The expression of $\hat V_v^{-\frac12}$ reads
\begin{equation}
\begin{aligned}
&\hat V_v^{-\frac12}=-\frac{16}{3}\left(\frac{1}{i\ell_P^2 \beta}\right)^3\sum_{e(I),e(J),e(K) \text{ at } v} \epsilon^{IJK}\tr\left([\hat{h}_{e(I)},\hat V_v^{1/2}]\hat{h}_{e(I)}^{-1}[\hat{h}_{e(J)},\hat{V}_v^{1/2}]\hat{h}_{e(J)}^{-1}[\hat{h}_{e(K)},\hat{V}_v^{1/2}]h_{e(K)}^{-1}\right).
\end{aligned}
\end{equation}
Semiclassically $\hat V_v^{-\frac12}$ is not always positive, and the sign relates to the orientation $\mathrm{sgn}(\det(e))$. However, the sign is canceled in the classical limit of $M_{v,v'}$, since every term contains a pair of $\hat V_v^{-\frac12}$. %\textcolor{red}{$\mathrm{sgn}(\det(e))$}.

\section{B. Initial state}\label{Initial state}

Given the cubic graph $\ga_n$, Thiemann's coherent state $|\psi_{g(\ga_n)}\rangle$ is peaked at the phase space point $g(\ga_n)\equiv\{h_e,p_e^a\}_{e\subset \ga_n}$ \cite{thiemann2001gaugeII,thiemann2001gaugeIII}. $h_e$ is the SU(2) holonomy of Ashtekar-Barbero connection $\G^a_I+\b K^a_I$ ($\G$ is the spin connection and $K$ is the extrinsic curvature), and $p^a_e$ smears the densitized triad $\sqrt{\det(q)}\bm{e}^I_a$ on the 2-face dual to $e$. The expectation values $\langle\psi_{g(\ga_n)}|\hat{h}_{e}|\psi_{g(\ga_n)}\rangle=h_e+O(\ell_P^2)$ and $\langle\psi_{g(\ga_n)}|\hat{p}^a_{e}|\psi_{g(\ga_n)}\rangle=p_e^a+O(\ell_P^2)$ endow $\ga_n$ with semiclassical internal and external geometries. $|\O\rangle\in\ch$ relates to $|\psi_{\ga_n}\rangle\equiv|\psi_{g_{\rm flat}(\ga_n)}\rangle$ (up to normalization) peeked at the flat geometry $g_{\rm flat}(\ga_n)=\{h_{e(I)}=\mathbf{1}_{2\times 2},\,p_{e(I)}^a=p_1\delta^a_I \}_{e(I)}$ where $e(I)$ denotes the edge along $I$-direction ($I=1,2,3$). $|\psi_{\ga_n}\rangle$ endow $\ga_n$ with the vanishing extrinsic curvature and the flat lattice geometry with constant lattice spacing $a_n=\sqrt{p_n}$. The effective fermion Hamiltonian $\hat{ H}_{\ga_n}^F(\psi_{\ga_n})$ equals the standard fermion Hamiltonian $\hat{\bm H}_{\ga_n}^F$ on the flat lattice plus $O(\ell_P^2)$ corrections, because $\langle\psi_{\ga_n}| M_{v,v'}(\hat{h},\hat{p})|\psi_{\ga_n}\rangle\|\psi_{\ga_n}\|^{-2}$ equals to $M_{v,v'}({h},{p})$ evaluated at ${h}_e=\mathbf{1}_{2\times2},\ {p}_{e(I)}^a=p_1\delta^a_I$ up to $O(\ell_P^2)$ \cite{giesel2007algebraic}. Namely, 
\bee
\hat{ H}_{\ga_n}^F(\psi_{\ga_n})=\hat{\bm H}_{\ga_n}^F\lt[1+O(\ell_P^2)\rt],\quad \hat{\bm H}_{\ga_n}^F=\frac{i\hbar}{2a_n}\sum_{v\in\gamma_n}\sum_{I=1}^{3}\left(\hat{\zeta}_{v}^{\dagger}{\bm\sigma}^{I}\hat{\zeta}_{v+\delta_{I}}-\hat{\zeta}_{v+\delta_{I}}^{\dagger}{\bm \sigma}^{I}\hat{\zeta}_{v}\right).
\eee
It is standard to diagonalize $\hat{\bm H}_{\ga_n}^F$ by Fourier transformation $\hat{\zeta}_{\vec{k}}=\sqrt{\frac{1}{N_n^3}}\sum_{v\in \gamma_n}\hat{\zeta}_{v}e^{-ia_n\vec{k}\cdot\vec{v}}$ ($\vec{v}\in\mathbb{Z}^3$ and $\vec{v}=\vec{v}+N_n$):
\bee
\hat{\bm H}_{\ga_n}^F=\frac{-\hbar}{a_n}\sum_{k\in \mathrm{FBZ}(\ga_n)}\hat{\zeta}_{k}^{\dagger}\lt[\sum_{I=1}^{3}\sin(a_n k_{I})\bm{\sigma}^{I}\rt]\hat{\zeta}_{k}=\frac{-\hbar}{a_n}\sum_{k\in \mathrm{FBZ}(\ga_n)}\hat{\zeta}_{k}^{\dagger}\Theta(k)^{\dagger}\left(\begin{array}{cc}
-\mathfrak{s}({k}) & 0\\
0 & \mathfrak{s}({k})
\end{array}\right)\Theta({k})\hat{\zeta}_{k}
\eee 
where $\mathfrak{s}({k})=\sqrt{\sum_{I=1}^{3} \sin ^{2}(a_n k_{I})}$, and $\Theta(k)$ is the unitary matrix diagonalizing $\sum_{I=1}^{3}\sin(a_n k_{I})\bm{\sigma}^{I}$. The ground state $|\o_{\ga_n}\rangle$ associates to the negative zero-point energy $ -\ce_0\equiv -\frac{\hbar}{a_n}\sum_{k\in \mathrm{FBZ}(\ga_n)}\mathfrak{s}({k})[1+O(\ell_P^2)]$.

The expectation value of physical Hamiltonian $\hat{\bf H}$ at $|\psi_{\ga_n}\rangle\otimes |\o_{\ga_n}\rangle$ is negative and equals $-\ce_0$. The existence of zero-point energy is because the fermion Hamiltonian $H_\ga^F$ in $\hat{\bf H}$ is not normal ordered. It does not make sense to normal order $H_\ga^F$ given that $\hat{\bf H}$ is background independent. Interestingly, we have the semiclassical relation ${\bf H}+E_{\rm dust}=0$ where $E_{\rm dust}$ is the energy of {Gaussian} dust \cite{Giesel:2012rb,Kuchar:1990vy}. This relation is one of the starting point of the reduced phase space quantization. Viewing ${\bf H}$ as the expectation value determines $E_{\rm dust}=\ce_0$. Then $\langle\psi_{\ga_n}|\hat{\bf H}+E_{\rm dust}|\psi_{\ga_n}\rangle=\hat{ H}_{\ga_n}^F(\psi_{\ga_n})+E_{\rm dust}$ cancels the fermion zero-point energy and gives the normal ordered Hamiltonian on the flat spacetime up to $O(\ell_P^2)$.

\section{C. Semiclassical analysis}\label{app:semiclassical}

%$\delta\tau_j$ is  $\tau^{(j+1)}-\tau^{(j)}$, and $\eta_{i,j}(e)$, for $j\geq 0$, is given by $\tr(g_{i}^\dagger(e)g_j(e))=2\cosh(\eta_{i,j}(e))$ and, for $j=-1$, is given $\tr(g_{i}^\dagger(e)(u\cdot g_{-1})(e))=2\cosh(\eta_{i,-1}(e))$, with $u\cdot g_{-1}$ denoting the gauge  transformed value of $g_{-1}$. Note that, the definition of $\delta\tau_j$ means $\delta\tau_j>0$ for $\tau_j< \tau_1$ and   $\delta\tau_j<0$ for $\tau_j\geq \tau_1$.

%Here by $\tau^{(i)}\mapsto (g_i(\gamma),\zeta_i(\gamma))$, we mean the coherent state $|\psi_{g_i(\gamma)}\rangle\otimes |\phi_{\zeta_i(\gamma)}\rangle$ is passed at time $\tau^{(i)}$. 
%Let $j_1$ and $j_2$ correspond to $\tau_1$ and $\tau_2$ (see Fig. \ref{fig:paths})
%, i.e., $\tau_k=\tau^{(j_k)}$ with $k=1,2$. 
%Following the standard derivation, we insert the resolution of identity $\int\dd Z_j(\gamma)|Z_j(\gamma)\rangle\langle Z_j(\gamma)|$, to get 

\begin{figure}
    \centering
    \includegraphics[width=0.6\textwidth]{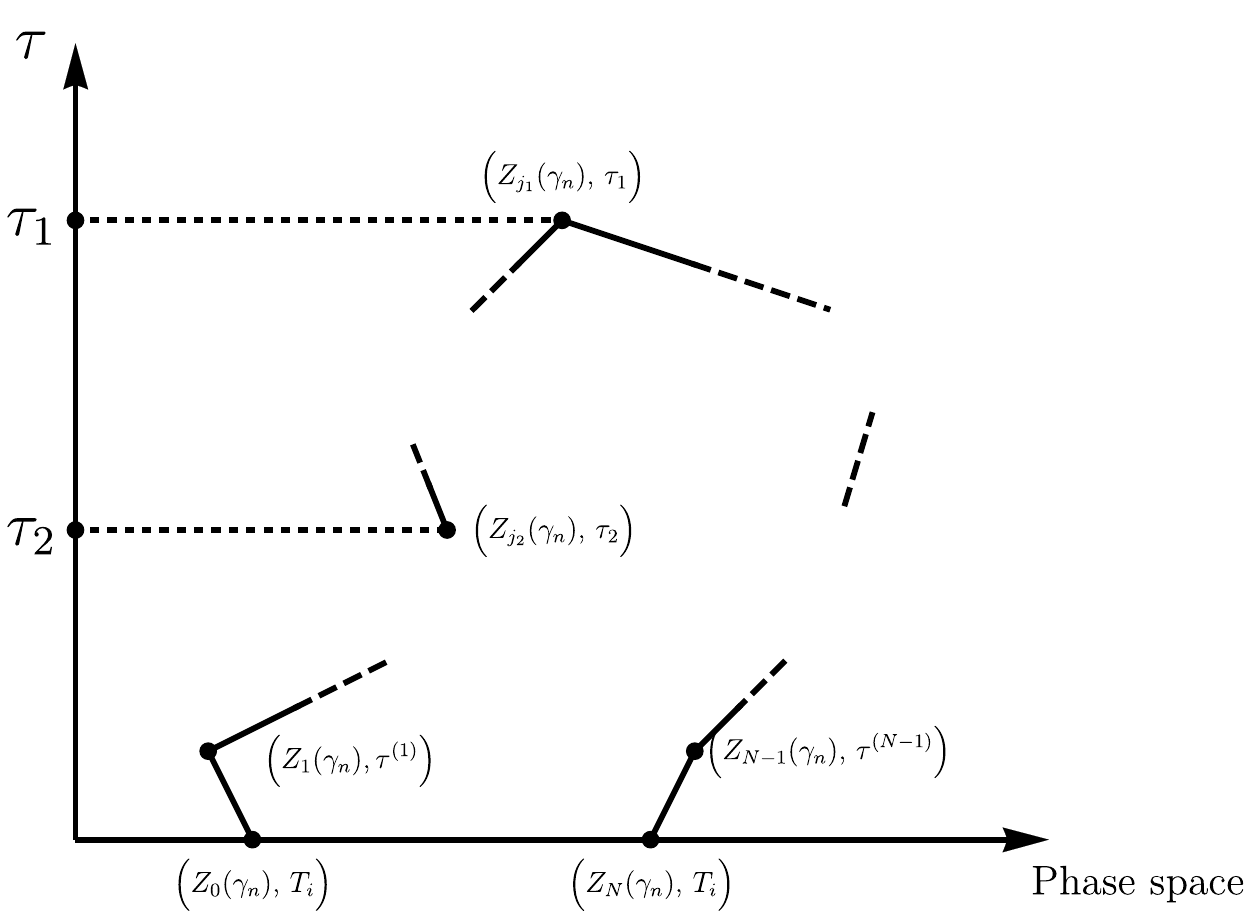}
    \caption{A $N$-side polygon path starting  from $T_i$, passing $\tau_2$, arriving at $\tau_1$ and returning $T_i$. $|Z_0(\gamma_n)\rangle$ and $|Z_N(\gamma_n)\rangle$ take the inner products with $|\O_{\gamma_n}\rangle$ and $|\O'_{\gamma_n}\rangle$ in \eqref{eq:gh} and \eqref{pathintegral00}.}
    \label{fig:paths}
\end{figure}

   %Note that \eqref{eq:gh} takes the inner products of the states $|\psi_{g_m(\gamma_n)}\rangle\otimes|\phi_{\zeta_m(\gamma_n)}\rangle$ for $m=0,N$ and the initial and final state.}

Let us consider the gauge invariant observable $\mathscr{G}_h(\tau_1,\tau_2,v_1,v_2)$. Similar to \eqref{GAB0}, we have
\bee
\mathscr{G}_h(\t_1,\t_2;v_1,v_2)&=&%\frac{1}{L}
\sum_{n=1}^L|w_n|^2\mathscr{G}_{h,n}(\t_1,\t_2;v_1,v_2),\\
\mathscr{G}_{h,n}(\t_1,\t_2;v_1,v_2)&=&\langle\O_{\ga_n}|(\hat{h}_c)_{BA}\ct[\hat{\Psi}^A(\t_1,v_1)\hat{\Psi}^B{}^\dagger(\t_2,v_2)]|\O_{\ga_n}\rangle.
\eee
We can obtain the path integral formula of $\mathscr{G}_{h,n}(\t_1,\t_2;v_1,v_2)$ on $\gamma_n$ as
\begin{equation}\label{eq:gh}
\begin{aligned}
\mathscr{G}_{h,n}(\tau_1,v_1,\tau_2,v_2)&=\int\limits_{\cg}\dd u\int\prod_{j=0}^{N}\rmd g_j(\ga_n)e^{S^{G}_{\ga_n}(g,u)}((\hat h_c)_{BA})_N(V^{-\frac{1}{2}}_{v_1})_{j_1}(V^{-\frac{1}{2}}_{v_2})_{j_2} \ci^{AB}_n(g,u)\\
\ci^{AB}_n(g,u)&=\int \prod_{j=0}^N \dd \zeta_j(\gamma_n)\, e^{S^F_{\gamma_n}(g,\zeta)}(\zeta_{v_1}^A)_{j_1}(\zeta_{v_2}^{B*})_{j_2}\langle\omega_{\ga_n}|\phi_{\zeta_N}\rangle\langle \phi_{\zeta_0}|u|\omega_{\ga_n}\rangle
\end{aligned}
\end{equation}
where the integral with respect $u=\{u_v\}_{v}\in\cg$ comes from the  projection $\p_{\mathcal G}$ introduced in defining $|\Omega_{\gamma_n}\rangle$. A pair of $\langle\O_{\gamma_n}$ and $|\O_{\gamma_n}\rangle$ only give one $u$-integral because the operator is gauge invariant and $\hat{\mathbf{P}}_{\cg}^2=\hat{\mathbf{P}}_{\cg}$.

% and $I^{AB}[p,h]$ in \eqref{gintegral} is the fermion path integral under the background corresponding to $(p,h)=\{g_i(\gamma)\}_{i=1}^N$, i.e.,$I^{AB}(g,u)=\int \prod_{i=0}^N \dd \zeta_i e^{S_F^\gamma(g,\zeta)}\zeta_{j_1,v_1}^A\zeta_{j_2,v_2}^{B*}\langle\omega_\gamma|\phi_{\zeta_N}\rangle\langle \phi_{\zeta_0}|u|\omega_\gamma\rangle$.  

In the derivation of the path integral, we insert the Heisenberg operator in $\mathscr{G}_h$, and have the following expression in terms of the time-evolution operators
\bee
\mathscr{G}_{h,n}(\tau_1,v_1,\tau_2,v_2)=\langle\O_{\ga_n}|(\hat{h}_c)_{BA}e^{\frac{i}{\hbar}\hat{\bf H}_{\ga_n}(\t_1-T_i)}\hat{\Psi}^A(v) e^{-\frac{i}{\hbar}\hat{\bf H}_{\ga_n}(\t_1-\t_2)}\hat{\Psi}^{B\dagger}(v) e^{-\frac{i}{\hbar}\hat{\bf H}_{\ga_n}(\t_2-T_i)}|\O_{\ga_n}\rangle,
\eee
when $\t_1>\tau_2$. It leads to that in the path integral \eqref{eq:gh}, the paths $
Z_j(\gamma_n)\equiv  (g_j(\gamma_n),\zeta_j(\gamma_n))$ start from $T_i$, pass $\tau_2$, arrive at $\tau_1$, and finally return to $T_i$. As in the standard derivation, we begin with considering the $N$-side polygon paths as shown in Fig. \ref{fig:paths} and, then, let $N$ approaches to $\infty$. In \eqref{eq:gh} and \eqref{gintegral}, $S_{\gamma_n}^G$ reads
\begin{equation*}
\begin{aligned}
S_{\gamma_n}^G=&\sum_{j=-1}^{N}\sum_{e\subset\gamma_n}\frac{2\eta_{j+1,j}(e)^2-\eta_{j+1,j+1}(e)^2-\eta_{j,j}(e)^2}{2\ell_P^2}-\frac{i }{\hbar} \sum_{\substack{j=0}}^{N-1}\rho_j\delta \tau\left( \frac{\langle  \psi_{g_{j+1}(\gamma_n)}|\p_{\gamma_n} \hat H_{\gamma_n}^G\p_{\gamma_n} |\psi_{g_j(\gamma_n)}\rangle}{\langle \psi_{g_{j+1}(\gamma_n)}| \psi_{g_j(\gamma_n)}\rangle})\right),
\end{aligned}
\end{equation*}
where $\rho_j=1$ or $-1$ at instances before or after $\t_1$. $j=-1$ and $j=N+1$ correspond to the initial and final state. The coherent label $g_j(\gamma_n)=\{g_e\}_{e\subset \gamma_n}$ relates to $h_e,p_e^a$ by $g_e=e^{-p_e^a\bm{\sig}^a/2}h_e$. $\eta_{i,j}(e)$ satisfies $\tr(g_{i}^\dagger(e)g_j(e))=2\cosh(\eta_{i,j}(e))$. At $j=-1$, $g_{-1}(\gamma_n)=u\cdot g_{\rm flat}(\gamma_n)$ is the gauge transformation of $g_{\rm flat}(\gamma_n)$.

$\ci^{AB}_n$ is independent of $\hbar$. The integral of $\dd u$ and $\dd g_j(\gamma_n)$ is studied with the stationary phase approximation to obtain a semiclassical expansion in $\ell_P^2$. With the chosen initial and final states, the solution of equations of motion gives $u=\mathring{u}=\mathbf{1}$ for all vertices $v$ and $g=\mathring{g}(\gamma_n)=(\mathring{p},\mathring{h})$ corresponding to the flat spacetime: $\mathring{h}_{e,j}=\mathbf{1}_{2\times 2}$ and $\mathring{p}_{e(I),j}^a=a_n^2\delta^a_I$. Thus as far as the leading order in the semiclassical expansion is concerned, we can just freeze $u$ in $\ci_n^{AB}$ to $\mathbf 1$, and freeze $((\hat h_c)_{BA})_N(V^{-\frac{1}{2}}_{v_1})_{j_1}(V^{-\frac{1}{2}}_{v_2})_{j_2}$ to their corresponding classical value at the flat geometry. Therefore,  Eq.\eqref{eq:gh} can be approximated by $\delta_{BA}\mathscr G^{AB}_n$ semiclassically, where $\mathscr G^{AB}_n$ is given by 
$$
\mathscr{G}^{AB}_n=\frac{1}{V_{\rm tot}}\ci^{AB}_{n}(\mathring{g},\mathring{u}) \int\limits_{\cg}\rmd u\int\prod_{j=0}^{N}\rmd g_j(\ga_n)\,e^{S^{G}_{\ga_n}(g,u)}\lt[1+O(\ell_P^2)\rt]
$$
The integral $\int\rmd u\int\prod_{j=0}^{N}\rmd g_j(\ga_n)\,e^{S^{G}_{\ga_n}(g,u)}$ is the transition amplitude between a pair of $|\tilde{\psi}_{\gamma_n}\rangle\equiv\p_\cg\p_{\gamma_n}|\psi_{\gamma_n}\rangle$, the quantum-geometry sectors of the initial state, with respect to the pure gravity Hamiltonian $\hat {\fh}_{\gamma_n}^G\equiv \p_{\gamma_n} \hat H_{\gamma_n}^G\p_{\gamma_n}$. Recall that the paths are the ones as in FIG.\ref{fig:paths}. We have 
\bee
\int\limits_{\cg}\rmd u\int\prod_{j=0}^{N}\rmd g_j(\ga_n)\,e^{S^{G}_{\ga_n}(g,u)}=\langle\tilde{\psi}_{\gamma_n}|e^{\frac{i}{\hbar}\hat{\fh}^G_{\gamma_n}(\t_1-T_i)} e^{-\frac{i}{\hbar}\hat{\fh}^G_{\gamma_n}(\t_1-\t_2)} e^{-\frac{i}{\hbar}\hat{\fh}^G_{\gamma_n}(\t_2-T_i)}|\tilde{\psi}_{\gamma_n}\rangle=\langle\tilde{\psi}_{\gamma_n}|\tilde{\psi}_{\gamma_n}\rangle=1+O(\ell_P^2).\nonumber
\eee
where in the last step, we use $1=\langle\O_{\gamma_n}|\O_{\gamma_n}\rangle=\int_{\cg} \rmd u \langle \psi_{\gamma_n}|\mathbf{P}_{\gamma_n}u\mathbf{P}_{\gamma_n}|\psi_{\gamma_n}\rangle\langle\o_{\gamma}|u|\o_{\gamma}\rangle=\int \rmd u \langle \psi_{\gamma_n}|\mathbf{P}_{\gamma_n}u\mathbf{P}_{\gamma_n}|\psi_{\gamma_n}\rangle[1+O(\ell_P^2)]$, because $\|\o_{\gamma_n}\|=1$ and $\langle \psi_{\gamma_n}|\mathbf{P}_{\gamma_n}u\mathbf{P}_{\gamma_n}|\psi_{\gamma_n}\rangle$ is a Gaussian-like function peaked at $u=\mathbf{1}$. Therefore
$$
\mathscr{G}^{AB}_n=\frac{1}{V_{\rm tot}}\ci^{AB}_{n}(\mathring{g},\mathring{u})[1+O(\ell_P^2)]. 
$$ 
Substituting the flat-geometry data $\mathring{p}$ and $\mathring{h}$ into $S^F_{\gamma_n}$, and performing the Fourier transformation to $\zeta_{i,v}$ along spatial directions, we get $S_{\gamma_n}^F$ in terms of the Fourier coefficients $\zeta_{j,\vec k}$ as
$S_N^F
= \sum_{j,\vec k,j',\vec k'}
\zeta_{j,\vec k}^\dagger\mathscr H_{j,\vec k,j',\vec k'} \zeta_{j',\vec k'}
$,
 with
$\mathscr H_{j,\vec k,j',\vec k'}= -\delta_{j,j'}\mathbf{1}_{2\times 2}+\delta_{j,j'+1}\left(\mathbf{1}_{2\times 2}+\frac{i \delta \tau}{a_n}  \sum_{I=1}^3 \sin(a_n k_I) {\bm \sigma}^I\right)$. With these results, $\ci^{AB}_{n}(\mathring{g},\mathring{u})$ can be computed by using the standard Fermionic Gaussian integral. The result is obtained by taking the limit $N\to\infty$: 
\bee
\ci^{AB}_{n}(\mathring{g},\mathring{u})=\sum_{\vec{k}\in \text{FBZ}(\ga_n) }\int_{\mathbb{R}}\frac{\rmd\omega}{2\pi i}\, e^{-i\omega(\tau_1-\t_2)+ia_n \vec{k}\cdot\left(\vec{v}_{1}-\vec{v}_{2}\right)}\,G_{\omega,\vec{k}}(\ga_n,a_n).
\eee

% \section{test plots}
% \begin{figure}[h!]
%	\begin{center}
%	\includegraphics[width=0.5\textwidth]{propagator_log.pdf}
%	\end{center}
%\end{figure}
% \begin{figure}[h!]
%	\begin{center}
%	\includegraphics[width=0.5\textwidth]{propagator_loglog.pdf}
%	\end{center}
%\end{figure}
 %\begin{figure}[h!]
%	\begin{center}
%	\includegraphics[width=0.5\textwidth]{plot1_lattice_refine_1.pdf}
%	\end{center}
%\end{figure}
 %\begin{figure}[h!]
%	\begin{center}
%	\includegraphics[width=0.8\textwidth]{plot1_lattice_refine.pdf}
%	\end{center}
%\end{figure}

\bibliography{reference}

\bibliographystyle{apsrev4-1}

\end{document}